# THE MASSES OF NEUTRON STARS

J.E. Horvath

Instituto de Astronomia, Geofísica e Ciências Atmosféricas USP, Departmento de Astronomia

Rua do Matão, 1226

05508-100 Cidade Universitária São Paulo, SP, Brazil

foton@iag.usp.br

R. Valentim

Departamento de Ciências Exatas e da Terra, Universidade Federal de São Paulo (UNIFESP)

Unidade José Alencar - Rua São Nicolau No 210

09913-030 Diadema, SP, Brazil

valentim.rodolfo@unifesp.br

## Abstract

We present in this article an overview of the problem of neutron star masses. After a brief appraisal of the methods employed to determine the masses of neutron stars in binary systems, the existing sample of measured masses is presented, with a highlight on some very well-determined cases. We discuss the analysis made to uncover the underlying distribution and a few robust results that stand out from them. The issues related to some particular groups of neutron stars originated from different channels of stellar evolution are shown. Our conclusions are that last century's paradigm that there a single, 1.4 $M_\odot$ scale is too simple. A bimodal or even more complex distribution is actually present. It is confirmed that some neutron stars have masses of $\sim 2\ M_\odot$, and, while there is still no firm conclusion on the maximum and minimum values produced in nature, the field has entered a mature stage in which all these and related questions can soon be given an answer.

## Introduction

Shortly after the discovery of pulsars (Hewish *et al.* 1968) and the identification of the pulsar in the Crab nebula (Staelin and Reifenstein 1968, see Sections 1.1 and 2.4), a great deal of theoretical and observational effort was directed to assess the physical features of these objects and their relation to their birth events. Even though early studies considered a wide range of possible values for the masses (i.e. theoretical cooling (Tsuruta et al. 1972) and related works (Sections 7.2 and 7.7)), the work on the pre-supernova evolution provided a clue about a possible "canonical" 1.4 $M_\odot$ imprinted on neutron stars at birth. This value was justified by the state of an iron core in a massive star just prior to the beginning of its collapse (Section 4.3). The mass of this core is (almost) invariant, since it has to be supported by electron degeneracy pressure. This core turns into a neutron star (after radiating ~10% of its energy) of a slightly lower mass. Later observational work was actually very successful in reducing the errors (see below), thus providing support to the idea of a single-mass scale.

Moreover, for almost 30 years the measurements of available neutron star systems (Thorsett and Chakrabarty 1999) proved to be consistent with the postulated unique value as suggested by that theoretical idea. The actual accretion history of those systems did not seem to make a large difference on the final neutron star mass, at least at a first glance, and the 21th century started with this as a firm paradigm.

However, intensive work performed by independent groups, both in the field of theory and especially on the observational side, changed the situation and added considerable interest to the study of neutron star masses.

We shall present below the basic arguments and evidence leading to believe that the old paradigm of a single-mass scale has to be abandoned. First, the main tools and methods for the measurements of neutron stars masses will be presented. An evaluation of the distribution of masses and its theoretical context will follow. The chapter ends with the statement of the conclusions on this subject.

# 1 Methods and tools for the measurements of neutron star masses

## 1.1 Kepler's Third Law and the mass function

Presently all the determinations of nutron star masses have been performed for objects in binary systems. The fundamental quantity to be measured as a first step is the *mass function* of the binary

$$f = \frac{(m_2 \sin i)^3}{(m_1+m_2)^2} = \frac{4\pi^2 x^3}{T_\odot P_b^2} \tag{1}$$

which can be determined directly from the observables of the right-hand side, the projected semi-axis of the orbit ($x$) and the period of the binary ($P_b$) after the introduction of the constant $T_\odot = GM_\odot/c^3 = 4.925490947\ \mu s$. In addition to $P_b$ and $x$, three Keplerian parameters can also be measured (Manchester and Taylor 1977), namely the eccentricity $e$, and the time and longitude of the periastron $T_0$ and $\omega_0$. The complete determination of the system still requires the measurements of at least two post-Keplerian parameters which are (different) functions of the five Keplerian parameters. These post-Keplerian parameters are the advance of the periastron $\dot\omega$, the orbital decay of the period $\dot P_b$ (dominated by the emission of gravitational waves from the varying quadrupole moment along the orbit evolution), the $\gamma$ parameter combining the variations of the transverse Doppler shift and gravitational redshift and the so-called "range" $r$ and "shape" $s$. Since all them depend on the theory of gravitation, their measurement open the possibility of testing GR itself through pulsar timing. Even if only one post-Keplerian parameter is measured, some statements about the masses are possible. However, these measurements are possible with accuracy only for close binaries with eccentric orbits. Therefore, in the general case, additional information on the companion mass $m_2$ and the inclination of the orbit respect to the line of sight $\sin i$ should be provided by independent techniques to determine the neutron star mass $m_1$.

## 1.2 Photometry, spectroscopy and related complementary tools

The theory of stellar evolution and the tools developed by astronomers along the 20th century are key ingredients to obtain complementary information to determine the masses of neutron stars. The systems in which the companion is directly observed feature main sequence, post-main sequence and white dwarf stars. Optical observations can be performed, but they are of little utility whenever $m_1 \ll m_2$, which is the case for example of the PSR J0045-7319 system (Nice et al. 2004) in which large errors remain because of the limited usefulness of the mass function in eq.(1). Low-mass companions and evolved objects for which $m_1 \approx m_2$ are more amenable to optical observations, in the sense of the complementary information needed to determine the neutron star mass $m_1$. The case of white dwarfs is particularly relevant, since a few of the most interesting systems belong to this class. Even though the white dwarfs can be very faint ($V \geq 24$ is not uncommon), their radii can be nevertheless estimated by measuring the optical flux $F_O$, the effective temperature $T_{eff}$ and having a good idea of the distance $d$, through the simple relation

$$R_{WD} = \left(\frac{F_O}{\sigma}\right)^{1/2} \left(\frac{d}{T_{eff}^2}\right) \tag{2}$$

where $\sigma$ is the Stefan-Boltzmann constant. Once the radius is determined, a reasonable estimate of the white dwarf mass can be made using the theoretical relations for a given composition, with temperature corrections to the cold equation of state if necessary. An alternative approach is to fit a synthetic atmosphere to the observed spectral features, in order to obtain the surface gravity $\log g$. The latter quantity combined with $T_{eff}$ for example, gives a determination of the mass, which of course subject to some uncertainty coming from the effects of finite temperature and the chemical abundances. The masses of $m_2$ may be overestimated and this effect is translated to the pulsar mass $m_1$ in the final calculations.

Spectral information of the system, in particular variability along the orbit, can be of great importance especially in the cases of interacting systems. The latter generally display variability due to accretion/winds which should be understood properly before any statement about the mass of the neutron star can be made. Thus, additional models for the geometry of the system should be used, and in extreme cases the heating of the companion by the winds are important and may dominate the uncertainties (see below). It is unfortunate that one particular feature, namely the presence of redshifted lines from the neutron star itself, does not seem to be present in the X-ray spectra with sufficient significance. If measured, redshifted lines carrying information about the quotient $M/R$ could be very useful. For example, the claim by Cottam et al. (2002) on the presence of absorption lines in the X-ray spectra bursts of the source EXO0748-676 could not be confirmed, but constitute an excellent example of how to extract important information about a compact star which may be attempted in other cases.

An important complementary method to determine pulsar masses is the Shapiro delay, i.e. the retardation of the pulses due to the gravitational field of the companion. The effect is particularly strong when nearly edge-on binaries are observed. The two parameters that control the behavior of the pulses are then $m_2$ (generating the field) and $\sin i$. The measurement of the arrival times may be combined with the Keplerian parameters to yield the masses and the inclination which reproduce the data best.

Finally a handful of alternative methods including polarization measurements (Thorsett and Stinebring 1990); scintillation (Cordes 1986) and even microlensing of background stars by *isolated* pulsars (Horvath 1996; Dai et al. 2015) have been proposed but did not produce sensible results as yet. More work will be needed to convert these ideas into useful tools for pulsar mass determinations.

## 2 The sample and the analysis of the neutron star mass distribution

The largest sample of measured neutron star masses available for analysis is maintained by J. Lattimer The compilation is publicly accessible on-line at http://www.stellarcollapse.org/. It is periodically updated with new determinations to keep it useful for the community. The latest (Dec 2015) version was made available for this paper by its author. Systems are separated into four classes according to the nature of the companion, and include neutron star-neutron star (double neutron star, in which at least one component is a pulsar); neutron-star-white dwarf; neutron star-main sequence and neutron star-X-ray/optical binaries. The references in Fig. 1 describe in detail each of the works reporting the techniques and models employed for the determinations.

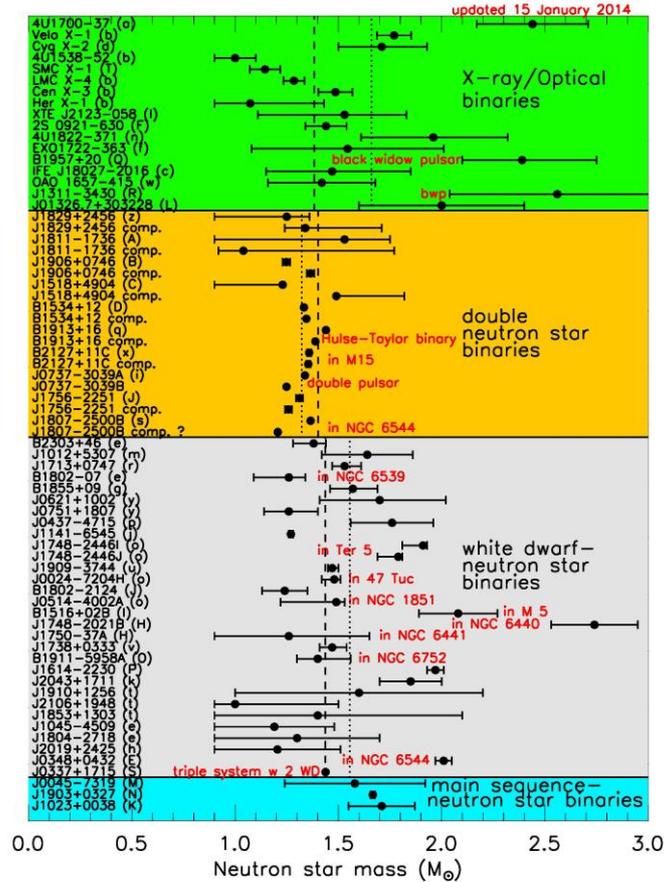

Fig. 1. The sample of neutron star masses maintained at stellarcollapse.org as displayed Oct. 28, 2015. The four groups are explicitly indicated, and the letters correspond to the references to the original works. The best determined values are accurate to the point in which the error bars fall inside the symbols. Systematic errors can shift the centroid in some cases, and constitute a major warning against a literal interpretation of the given numbers. References are indicated with a letter on each point: [a] Clark et al.(2002); [b] Rawls et al. (2011); [c] Mason et al.(2011); [d] Casares et al.(2010); [e] Thorsett and Chakrabarty (1999); [f] Mason et al.(2010); [g] Nice, Splaver and Stairs (2005); [h] Nice et al. (2001); [j] Bhat et al. (2008); [k] Guillemot et al. (2012); [l] Gelino et al. (2003); [m] Lange et al. (2001); [n] Muñoz-Darias et al. (2005); [o] Kiziltan et al. (2013); [p] Verbiest et al. (2008); [q] Weisberg et al. (2010); [r] Splaver et al. (2005); [s] Lynch et al. (2012); [t] González et al. (2011); [u] Hotan et al. (2006); [v] Antoniadis et al. (2012); [w] Mason et al. (2012); [x] Jacoby et al. (2006); [y] Nice et al. (2008); [z] Champion et al. (2005); [A] Corongiu et al. (2007); [B] Kasian (2008); [C] Janssen et al. (2008); [D] Stairs et al. (2002); [E] Antoniadis et al. (2013); [F] Steeghs and Jonker (2007); [H] Freire et al. (2008b); [I] Freire et al. (2008a); [J] Ferdman et al. (2010); [K] Deller et al. (2012); [L] Bhalerao et al. (2012); [M] Nice et al. (2004); [N] Freire et al. (2011); [O] Bassa et al. (2006); [P] Demorest et al. (2010); [Q] van Kerkwijk et al. (2011); [R] Romani et al. (2012); [S] Ransom et al. (2014); [T] Coe et al. (2013).

A few remarkable cases among these mass determinations merit a highlight. The first is the very high accuracy achieved after several decades of work in the double neutron star systems. In these cases, the determination of all post-Keplerian parameters has been possible and therefore the masses are now known beyond 4 decimal places in at least one system. This is a remarkable achievement and puts double neutron stars among the most accurately measured star masses overall, including their "normal" main sequence relatives. In fact, the latest determination of a double neutron star system by Martínez et al. (2015) yielded a very asymmetric system, in contrast with the nearly equal-mass binaries of this type known up today. In addition, the mass of one of the stars is only $m_2 = 1.174 \pm 0.004\, M_\odot$, which is the lowest mass measured with confidence. If it is indeed a neutron star, this determination contributes to the quest of the *minimum* mass that can be produced in nature, which is set by evolutionary considerations and not by the existence of any physical limit, at least in the measured range.

Another benchmark determination was the work of Demorest et al. (2010) measuring the Shapiro delay in the system of PSR J1614−2230. This effect stems from the effects of the gravitational well of the companion as is seen at different phases along the orbit. In general, and even after several years of timing, the Shapiro delay may remain "hidden" if it is too small. The case of the PSR J1614−2230 benefited from a combination of an almost edge-on system with a relatively high mass of the companion white dwarf ($0.500 \pm 0.006\, M_\odot$). The determination with high accuracy needed a sophisticated statistical analysis to subtract a full GR timing (i.e. non-Shapiro delay contributions) and provided a firm number for the mass as $m_1 = 1.97 \pm 0.04\, M_\odot$, now widely accepted by the community. This stands among the top masses ever measured with a value well beyond the "old" $1.4\, M_\odot$ paradigm.

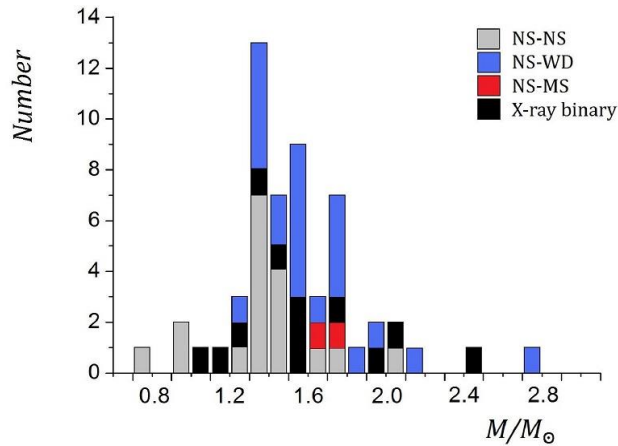

Fig. 2. The histogram of the masses of neutron stars in binary systems presented by Valentim et al. (2011). Note that new objects have been added to the database in Fig. 1, and therefore the exact numbers of the mass scales may differ when the updated database is used.

A similar number for the mass of the system PSR J0348+0432 has been obtained by Antoniadis et al. (2013) using the methods described above. The companion is a low-mass helium white dwarf which was characterized by a combination of phase-resolved spectra, fitting of synthetic spectra and a theoretical finite-temperature mass-radius relation. The fitting procedures yielded very accurate results (Fig. 3), and therefore the masses of the components were determined. The high value $m_1 = 2.01 \pm 0.04\, M_\odot$ reinforces the presence of a mass-scale substantially higher than the $1.4\, M_\odot$ formerly supported. Both $\sim 2\, M_\odot$ objects are therefore examples that the accretion history can substantially affect the mass of a compact object although the exact amount of matter onto it depends on several factors that vary in each case (Sections 7.13 and 7.14).

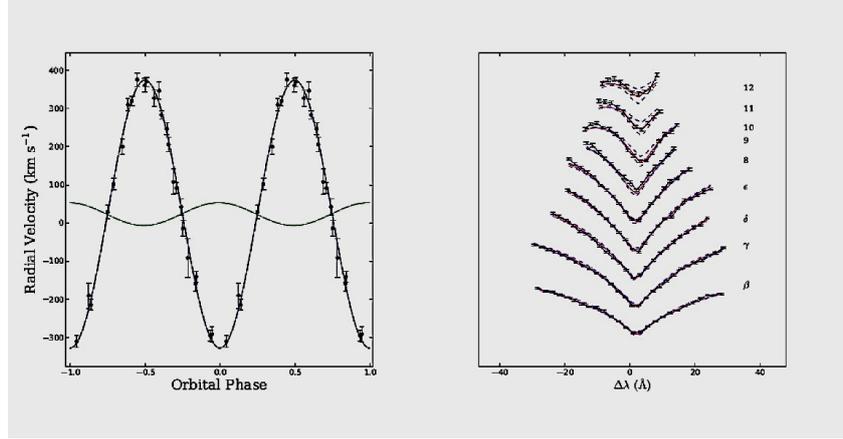

Fig. 3. The fittings to the radial velocity (left) and Balmer lines (right) obtained by Antoniadis et al. (2013) to the system harboring PSR J0348+0432 and its white dwarf companion. Solid lines are the best fit to the orbit (left) and three atmospheric models, showing that slight changes in $T_{eff}$ and/or $\log g$ worsen the accurate result obtained for the values $(10120 \pm 47_{stat} \pm 90_{sys}, 6.035 \pm 0.032_{stat} \pm 0.06_{sys})$. These measurements, when combined with the Keplerian parameters, determine the value of the pulsar with high accuracy.

With these considerations in mind, we shall now discuss the main features of the analysis performed and the meaning of the results.

## 2.1 Analysis of the neutron star mass distribution

At least four different groups (Zhang et al. 2011; Valentim et al. 2011; Özel et al. 2012; Kiziltan et al. 2013) have presented independent analysis of the mass distribution (Fig. 2). All but Zhang et al. (2011) employed Bayesian analysis techniques. While they differ somewhat in the criteria to select the size of the sample (for example, trying to avoid a contamination with biased/uncertain determinations), all the reported results are indicative of the presence of at least two mass-scales. The group of Özel et al. (2012) found a mean mass of $1.28\,M_\odot$ for non-recycled high-mass binaries and slow pulsars, $1.33\,M_\odot$ for double neutron stars and $1.48\,M_\odot$ for recycled neutron stars, all them showing different dispersions. Zhang et al. (2011) reported a bimodality at $1.37\,M_\odot$ and $1.57\,M_\odot$ when the sample is divided by a 20 ms period possibly separating the non-recycled to the recycled population, although their true focus was somewhat different than the other works, being more related to the spin evolution of the systems itself. Kiziltan et al. (2013), on the other hand, presented an analysis in which just double neutron stars and neutron stars with white dwarf companions were selected. Their results yielded peaks at $1.33\,M_\odot$ and $1.55\,M_\odot$ and allowed skewed distributions. The work of Valentim et al. (2011) included all the objects available in the stellarcollapse.org database at the time and found (within a gaussian parametrization as in Özel et al. (2012)) the values $1.37\,M_\odot$ and $1.73\,M_\odot$, also with quite different widths (narrow and wide respectively) for the assumed gaussian profiles. These differences may be entirely due to the selection of the sample itself.

With the availability of more measured masses these analysis can be refined and further compared. For example, the issue of the masses "at birth" complicates a clear separation between non-recycled and recycled systems from the point of view of the total accreted mass. This is because of the well-known jump in the iron core for progenitor masses $> 19\,M_\odot$ in the main sequence. The generated iron cores may indeed grow to $\geq 1.8\,M_\odot$ at the moment of collapse because of convection in the carbon burning stage

(Timmes et al. 1996). Therefore, and after allowing for the binding energy of the neutron star respect to the previous state, there may be a ~0.3 $M_\odot$ difference at birth between these heavy objects and the more common ~1.4 $M_\odot$ cousins coming from lighter progenitors. The degeneracy between the initial mass and accreted mass could be important in many cases and is not easy to break. In other words, even if the accretion has been substantial, the initial mass of the neutron star cannot be determined with precision, and the assumption of a 1.4 $M_\odot$ often made could be quite misleading.

## 2.2 Existence of a group of neutron stars around 1.25 $M_\odot$

Stellar evolution calculations agree on the formation of very degenerate O-Mg-Ne cores for the lowest end of stars $8 - 11\ M_\odot$ (the latter figure depending on the detailed physics of the evolution, which could change somewhat the actual number). These cores are expected to eventually collapse because of electron capture. Some actual systems have been discussed in Poelarends et al. (2008) among other works. Podsiadlowski et al. (2005) proposed that this "characteristic mass" of the neutron stars formed from this mass range, with a typical amount of ejected mass, would finally produce low-mass neutron stars of ~ 1.25 $M_\odot$. The formation of this "canonical" mass and its identification as a separate peak in the mass distribution suggests one robust evolutionary path towards collapse/formation and should occur, preferentially, in binaries with low eccentricity and aligned orbits (Schwab et al. 2008) to form the systems in which the masses are presently observed. In those binary systems, low masses are formed when a white dwarf has O-Ne-Mg core accretes mass from the companion. The core density reaches a well-defined critical value ($\sim 4.5 \times 10^9\ g\ cm^{-3}$) triggering electron captures onto Mg and subsequently Ne, and causing a loss of hydrostatic support in the core and the onset of the collapse. These are the key aspects of these *e-capture supernova*, and since the $8 - 11\ M_\odot$ (and possibly even up to 12 $M_\odot$, see Section 4.2) progenitor stars are the most abundant among massive stars, their neutron star descendants should be well represented in the sample.

An inspection of Fig. 1 reveals a handful of objects with masses consistent with this value. In fact, even if former analysis (Valentim et al. 2011) casted some doubts on the significance of the peak, newer evidence has reinforced the idea that e-capture supernova neutron stars are indeed present in the sample, as argued by Schwab et al. (2008). It is interesting to point out that the "old" picture of a single-mass scale had not identified any problem between the theoretical prediction (Timmes et al. 1996) and the actual measured values, which in spite of being close should have stood as a separate channel.

## 2.3 Heavy neutron stars in close binary systems ("spiders")

There is one class of interacting millisecond pulsars in close binaries which is particularly interesting to assess the effects of the accretion history and maximum achievable mass. The first object of this type was discovered by Frutcher et al. (1988), and showed signals that an outflow was evaporating the companion, now reduced to a Jupiter-scale object. Because of a previous accretion phase in which the pulsar become accelerated to millisecond periods, while its wind was destroying the donor star at present, this system was dubbed a *black widow* in parallel to the behavior of that class of spiders. Later, a similar group was identified in a different region of the orbital period-donor mass, and received the name of *redbacks* which are Australian spiders related to black widows (Roberts 2013). The connection between the two groups was explored by means of evolutionary calculations (Benvenuto et al. 2012), and it was shown that a long accretion stage ($\geq 2 - 3\ Gyr$) shaped by X-ray back illumination led to systems (within a very restricted region of parameter space) which transit towards the redback region, and later when the donor became degenerate, widen their orbits while the evaporation of the donor proceeded.

The importance of these systems for the problem of neutron star masses stems from the very long evolution times, and in spite that the exact amount of mass ending on the neutron star itself could not be calculated, several tenths of $M_\odot$ are expected as a general feature (see Tauris 2015 for an alternative view and connections with other neutron star systems). This is why the determinations of van Kerkwijk et al. (2011) for the "original" black widow PSR 1957+20 (2.4 $\pm$ 0.12 $M_\odot$) and Romani et al. (2012) in the case of PSR J1311-3430 (concluding that $m_1 > 2.1\ M_\odot$) came to support these theoretical trajectories. However, the issue of the masses is far from being settled. For example, Romani et al. (2015) found that a simple direct heating model formerly employed for the atmosphere of PSR J1311-3430 is inadequate and therefore a systematic deviation of the mass from its true value leads to an unreliable estimation. The mass of the pulsar in the black widow system PSR J1311-3430 may be as low as $1.8\ M_\odot$. A confirmation of the mass of PSR 1957+20 with confidence on the absence of possible systematic effects would be very important as well. In any case, the black widow and redback pulsars are expected to contain the most massive neutron stars in nature, and this is why their study is so important for this field (Section 7.6). The measurements of magnetic fields in the ball park of $10^8 G$ for systems with ages $\geq 10\ Gyr$ is also revealing features of magnetic field evolution (Section 7.4) which are still under work when this article goes to the press.

## Conclusions

We have presented the general outline of neutron star mass determinations and discussed the picture emerging from the analysis of the available sample, comprising more than 70 objects at present. For most of the determinations, the systematic errors still affect the determinations, and the observations carry error bars that are significant in most cases. Exceptions to this picture include the binary neutron stars, and a few other objects. The most important conclusion of the large amount of work performed in the field is that the "old" view of a single $\sim 1.4\ M_\odot$ mass scale is untenable, since evidence for substantially heavier neutron stars is now available. It is still unclear exactly what kind of distribution is present in the data, and while bimodality is rooted in the theoretical framework, the role of the accretion history of the systems has to be clarified to address this point. It is also fair to state that the peak at $\sim 1.25\ M_\odot$ expected to form from the lightest progenitor collapses in the range $8 - 11\ M_\odot$ is actually present in the sample with increasing levels of significance. Finally, we should point out that there is no hint as yet about the *actual* maximum mass of neutron stars, but only a consensus on the $\sim 2\ M_\odot$ determinations which seem robust. The upper limit may be set by evolution (Kiziltan et al. 2013) or fundamental physical factors, and is of course extremely important to address the nature of the equation of state above nuclear saturation density, and the nature of matter under extreme conditions (Lattimer 2012).

## Acknowledgements

The authors wish to acknowledge the financial support of the Fapesp Agency (São Paulo) through the grant 2013/26258-4. J.E.H. has been partially supported by the CNPq Agency (Brazil) by means of a Research Fellowship.

## References

Antoniadis J, van Kerkwijk MH, Koester D, Freire PCC, Wex N, Tauris TM, Kramer M, Bassa CG (2012) The relativistic pulsar-white dwarf binary PSR J1738+0333 - I. Mass determination and evolutionary history. MNRAS 423, 3316-3327 [v]

Antoniadis J, Freire PCC, Wex N, Tauris TM, Lynch RS, van Kerkwijk MH, Kramer M, Bassa C, Dhillon VS, Driebe T, Hessels JWT, Kaspi VM, Kondratiev VI, Langer N, Marsh TR, McLaughlin MA,

Pennucci TT, Ransom SM, Stairs IH, van Leeuwen J, Verbiest JPW, Whelan DG (2013) A Massive Pulsar in a Compact Relativistic Binary. Science 340: 448-450 [E]

Bassa CG, Jonker PG, in't Zand JJM, Verbunt F (2006) Two new candidate ultra-compact X-ray binaries. A&A 446: L17-L20 [O]

Bhat NDR, Bailes M, Verbiest JPW (2008) Gravitational-radiation losses from the pulsar white-dwarf binary PSR J1141 6545. Phys. Rev. D 77: 124017 [j]

Bhalerao VM, van Kerkwijk MH, Harrison FA (2012) Constraints on the Compact Object Mass in the Eclipsing High-mass X-Ray Binary XMMU J013236.7+303228 in M 33. ApJ 757: 10 [L]

Casares J, Gonzalez Hernandez JI, Israelian G, Rebolo R (2010) On the mass of the neutron star in Cyg X-2. MNRAS 401: 2517-2520 [d]

Champion DJ, Lorimer DR, McLaughlin MA, Xilouris KM, Arzoumanian Z, Freire PCC, Lommen AN, Cordes, JM, Camilo, F (2005) Arecibo timing and single-pulse observations of 17 pulsars. MNRAS 363: 929-936 [z]

Clark J., Goodwin SP, Crowther PA, Kaper L, Fairbairn M, Langer N, Brocksopp C (2002) Physical parameters of the high-mass X-ray binary 4U1700-37. A&A 392: 909-920 [a].

Coe MJ, Angus R, Orosz JA, Udalski, A (2013) A detailed study of the modulation of the optical light from Sk160/SMC X-1. MNRAS 433: 746-750 [T]

Cordes J (1986) Space velocities of radio pulsars from interstellar scintillations. ApJ 311: 183-196

Corongiu A, Kramer M, Stampers BW, Lyne AG, Jessner A, Possenti A, D'Amico N, Löhmer O (2007) The binary pulsar PSR J1811-1736: evidence of a low amplitude supernova kick. A&A 462: 703-709 [A]

Cottam J, Paerels F, Méndez M (2002) Gravitationally redshifted absorption lines in the X-ray burst spectra of a neutron star. Nature 420: 51-54

Dai S, Smith MC, Lin MX, Yue YL, Hobbs G, Xu RX (2015) Gravitational Microlensing by Neutron Stars and Radio Pulsars: Event Rates, Timescale Distributions, and Mass Measurements. ApJ 802: A120

Deller, A. T, Archibald AM, Brisken WF, Chatterjee S, Janssen GH, Kaspi VM, Lorimer D, Lyne AG, McLaughlin MA, Ransom S, Stairs IH, Stappers B (2012) A Parallax Distance and Mass Estimate for the Transitional Millisecond Pulsar System J1023+0038. ApJL 756: L25-L30 [K]

Demorest PB, Pennucci T, Ransom SM, Roberts MSE, Hessels JWT (2010) A two-solar-mass neutron star measured using Shapiro delay. Nature 467: 1081-1083 [P]

Ferdman RD (2008) Ph.D. thesis, Univ. of British Columbia [J]

Ferdman RD, Stairs IH, Kramer M, McLaughlin MA, Lorimer DR, Nice DJ, Manchester RN, Hobbs G, Lyne AG, Camilo F, Possenti A, Demorest PB, Cognard I, Desvignes G, Theureau G, Faulkner A, Backer DC (2010) A Precise Mass Measurement of the Intermediate-Mass Binary Pulsar PSR J1802 – 2124. ApJ 711: 764-771 [R]

Freire PCC, Wolszcan A, van den Berg M, Hessels JWT (2008a) A Massive Neutron Star in the Globular Cluster M5. ApJ 679: 1433-1442 [I]

Freire PCC, Ransom SM, Bégin S, Stairs IH, Hessels JWT, Frey LH, Camilo F (2008b) Eight New Millisecond Pulsars in NGC 6440 and NGC 6441. ApJ 675: 670-682 [H]

Freire PCC, Bassa CG, Wex N, Stairs IH, Champion DJ, Ransom SM, Lazarus P, Kaspi VM, Hessels JWT, Kramer M, Cordes JM, Verbiest JPW, Podsiadlowski P, Nice DJ, Deneva JS, Lorimer DR, Stappers BW, McLaughlin MA, Camilo F (2011) On the nature and evolution of the unique binary pulsar J1903+0327 MNRAS 412: 2763-2780 [N]

Frutcher, A.S., Stinebring, D.R., Taylor, J.H. Nature **333**, 237 (1988)

Gelino DM, Tomsick JA, Heindl WA (2002) Measuring the Orbital Inclination Angle for the Low-Mass X-Ray Binary XTE J2123-058. Bull. Am. Astron. Soc. 34: 1199 [l]

González ME, Stairs IH, Ferdman RD, Freire PCC, Nice DJ, Demorest PB, Ransom SM, Camilo F, Hobbs G, Manchester RN, Lyne AG (2011) High-precision Timing of Five Millisecond Pulsars: Space Velocities, Binary Evolution, and Equivalence Principles. ApJ 743: A102 [t]

Guillemot L, Freire PCC, Cognard I, Johnson,TJ, Takahashi Y, Kataoka J, Desvignes G, Camilo F, Ferrara EC, Harding AK, Janssen GH, Keith M, Kerr M, Kramer M, Parent D, Ransom SM, Ray PS, Saz Parkinson PM, Smith DA, Stappers BW, Theureau G (2012) Discovery of the millisecond pulsar PSR J2043+1711 in a Fermi source with the Nançay Radio Telescope. MNRAS 422: 1294-1305 [k]

Hewish A, Bell SJ, Pilkington JDH, Scott PF, Collins RA (1968) Observation of a Rapidly Pulsating Radio Source. Nature 217: 709-713

Horvath JE (1996) Possible determination of isolated pulsar masses with gravitational microlensing. MNRAS 278: L46-L48

Hotan AW, Bailes M, Ord SM (2006) High-precision baseband timing of 15 millisecond pulsars. MNRAS 369: 1502-1520 [u]

Jacoby, B.A, Cameron PB, Jenet FA, Anderson SB, Murty RN, Kulkarni SR (2006) Measurement of Orbital Decay in the Double Neutron Star Binary PSR B2127+11C. ApJL 644: L113-L116 [x]

Janssen GH, Stappers BW, Kramer M, Nice DJ, Jessner A, Cognard I, Purver MB (2008) Multi-telescope timing of PSR J1518+4904. A&A 490: 753-763 [C]

Kasian L (2008) Timing and Precession of the Young, Relativistic Binary Pulsar PSR J1906+0746. In: Bassa C, Wang Z, Cumming A, Kaspi VM (eds) AIP Conf. Ser. 983, 40 Years of Pulsars: Millisecond Pulsars, Magnetars and More, AIP, New York pp.485-487 [B]

Kiziltan B, Kottas A, De Yoreo M, Thorsett SE (2013) The Neutron Star Mass Distribution. ApJ 778: A66 [o]

Kramer, M, Stairs IH, Manchester RN, McLaughlin MA, Lyne AG, Ferdman RD, Burgay M, Lorimer DR, Possenti A, D'Amico N, Sarkisian JM, Hobbs GB, Reynolds JE, Freire PCC, Camilo F (2006) Tests of General Relativity from Timing the Double Pulsar. Science 314: 97-102 (2006) [i]

Lange Ch, Camilo F, Wex N, Kramer M, Backer DC, Lyne AG, Doroshenko O (2001) Precision timing measurements of PSR J1012+5307. MNRAS 326: 274-282 [m]

Lattimer J (2012) The Nuclear Equation of State and Neutron Star Masses. Annu. Rev. Nuc. Part. Phys. 62: 485-515

Lynch RS, Freire PCC, Ransom SM, Jacoby BA (2012) The Timing of Nine Globular Cluster Pulsars. ApJ 745: A109 [s]

Manchester RN, Taylor JH (1977) Pulsars. Freeman, San Francisco

Martínez JG, Stovall K, Freire PCC, Deneva JS, Jenet FA, McLaughlin MA, Bagchi M, Bates SD, Ridolfi A (2015) Pulsar J0453+1559: A Double Neutron Star System with a Large Mass Asymmetry. ApJ 812: A143

Mason AB, Norton AJ, Clark, JS, Negueruela I, Roche P (2010) Preliminary determinations of the masses of the neutron star and mass donor in the high mass X-ray binary system EXO 1722-363 . A&A 509: A79 [f]

Mason AB Norton AJ, Clark, JS, Negueruela I, Roche P (2011) The masses of the neutron and donor star in the high-mass X-ray binary IGR J18027-2016. A&A 532: A124 [c]

Mason AB, Clark JS, Norton AJ, Crowther PA, Tauris TM, Langer N, Negueruela I, Roche P (2012) The evolution and masses of the neutron star and donor star in the high mass X-ray binary OAO 1657-415. MNRAS 422: 199-206 [w]

Muñoz-Darias T, Casares J, Martínez-Pais IG (2005) The "K-Correction" for Irradiated Emission Lines in LMXBs: Evidence for a Massive Neutron Star in X1822-371 (V691 CrA). ApJ 635: 502-507 [n]

Nice DJ, Splaver EM, Stairs IH (2001) On the Mass and Inclination of the PSR J2019+2425 Binary System. ApJ 549: 516-521 [h]

Nice DJ Splaver EM, Stairs IH (2004) Heavy Neutron Stars? A Status Report on Arecibo Timing of Four Pulsar - White Dwarf Systems. In: Camilo F, Gaensler BM (eds) Young neutron stars and their environments, IAU Symposium 218, Astr. Soc.Pac., San Francisco , pp 282 (2004) [M]

Nice DJ, Splaver EM, Stairs IH (2005) Masses, Parallax, and Relativistic Timing of the PSR J1713+0747 Binary System. In: Rasio F, Stairs IH (eds) Binary Radio Pulsars, Astr. Soc.Pac., San Francisco 328, pp 371 [g]

Nice, D.J., Stairs, I.H., Kasian, L.E. (2008) Masses of Neutron Stars in Binary Pulsar Systems. In: Bassa C, Wang Z, Cumming A, Kaspi VM (eds) AIP Conf. Ser. 983, 40 Years of Pulsars: Millisecond Pulsars, Magnetars and More, AIP, New York pp.453-458 [y]

Özel F, Psaltis D, Narayan R, Santos Villareal A (2012) On the Mass Distribution and Birth Masses of Neutron Stars. ApJ **757**: A55

Podsiadlowski Ph, Dewi JDM, Lesaffre P, Miller JC, Newton WG, Stone JR (2005) The double pulsar J0737-3039: testing the neutron star equation of state. MNRAS 361: 1243-1249

Poelarends AJT, Herwig F, Langer N, Heger A (2008) The Supernova Channel of Super-AGB Stars. ApJ 675: 614-625

Ransom S, Stairs IH, Archibald AM, Hessels JWT , Kaplan DL, van Kerkwijk MH, Boyles J, Deller AT, Chatterjee S, Schechtman-Rook A, Berndsen A, Lynch RS, Lorimer DR, Karako-Argaman C, Kaspi VM, Kondratiev VI, McLaughlin MA, van Leeuwen J, Rosen R, Roberts MSE, Stovall K (2014) A millisecond pulsar in a stellar triple system. Nature 505: 520-524 [S]

Rawls ML, Orosz JA, McClintock JE, Torres MAP, Bailyn CD, Buxton MM (2011) Refined Neutron Star Mass Determinations for Six Eclipsing X-Ray Pulsar Binaries. ApJ 730: 25-36 [b]

Roberts MSE (2013) Surrounded by spiders! New black widows and redbacks in the Galactic field. In van Leewen J (ed). Neutron Stars and Pulsars: Challenges and Opportunities after 80 years. Proceeding of the IAU 291, Cambridge University Press, Cambridge, pp. 127-132

Romani RW, Filippenko AV, Silverman JM, Cenko S, Greiner J, Rau A, Elliott J, Pletsch HJ (2012) PSR J1311-3430: A Heavyweight Neutron Star with a Flyweight Helium Companion. ApJL 760: L36-L41 [R]

Romani RW, Filippenko AV, Cenko, SB (2015) A Spectroscopic Study of the Extreme Black Widow PSR J1311-3430. ApJ 804: A115

Schwab J, Podsiadwolski Ph, Rappaport S (2010) Further Evidence for the Bimodal Distribution of Neutron-star Masses. ApJ 719: 722-727

Splaver EM, Nice DJ, Stairs IH, Lommen AN, Backer DC (2005) Masses, Parallax, and Relativistic Timing of the PSR J1713+0747 Binary System. ApJ 620: 405-415 [r]

Staelin DH, Reifenstein EC III (1968) Pulsating Radio Sources near the Crab Nebula. Science 162: 1481-1483

Stairs IH, Thorsett SE,Taylor JH, Wolszczan A (2002) Studies of the Relativistic Binary Pulsar PSR B1534+12. I. Timing Analysis. ApJ 581: 501-508 [D]

Steeghs D, Jonker PG (2007) On the Mass of the Neutron Star in V395 Carinae/2S 0921-630. ApJL 669: L85-L88 [F]

Tauris T (2015) *Habilitation Thesis*, arXiv:1501.03882

Thorsett SE, Stinebring DR (1990) Polarimetry of millisecond pulsars. ApJ 361: 644-649

Thorsett SE, Chakrabarty D (1999) Neutron Star Mass Measurements. I. Radio Pulsars. ApJ 512: 288-299 [e]

Tsuruta S, Canuto V, Lodenquai J, Ruderman M (1972) Cooling of Pulsars. ApJ 176: 739-750

Valentim R, Rangel E, Horvath JE (2011) On the mass distribution of neutron stars. MNRAS 414: 1427-1431

van Kerkwijk MH, Breton R, Kulkarni SR (2011) Evidence for a Massive Neutron Star from a Radial-velocity Study of the Companion to the Black-widow Pulsar PSR B1957+20. ApJ 728: A95 [Q]

Verbiest JPW, Bailes M, van Straten W, Hobbs GB, Edwards RT, Manchester RN, Bhat NDR, Sarkissian JM, Jacoby BA, Kulkarni SR (2008)  Precision Timing of PSR J0437-4715 (2008) An Accurate Pulsar Distance, a High Pulsar Mass, and a Limit on the Variation of Newton's Gravitational Constant. ApJ 679: 675-680 [p]

Weisberg JM, Nice DJ, Taylor JH (2010) Timing Measurements of the Relativistic Binary Pulsar PSR B1913+16. ApJ 722: 1030-1034 [q]

Zhang CM,  Wang J, Zhao YH,  Yin HX, Song LM, Menezes DP, Wickramasinghe DT, Ferrario L, Chardonnet P (2011) Study of measured pulsar masses and their possible conclusions .A&A 527: A83